# Influence of selection criteria on the interpretation of rotational behaviour of wide-binary star systems

S A Cookson⋆

*Independent Researcher, Crawley RH10 3EX, UK*

## ABSTRACT

Binary star systems are expected to follow Newtonian dynamics similarly to planetary systems. However, reports have been made of wide binary systems with separations around 0.01 pc and larger, showing potential deviations from standard Newtonian motion. This phenomenon, suggestive of the flattening of galactic rotation curves, calls for closer inspection. This study presents an analysis of wide binary stars using data from *Gaia* Data Release 3 (DR3), a space-based astrometry mission funded by the European Space Agency. The study compares different choices of selection criteria to examine the nature of these apparent anomalous kinematics within the solar neighbourhood. The *Gaia* data set furnishes detailed parameters, including radial velocity, mass, age, and binary probability in addition to standard parameters. A custom Python tool named BYNARY facilitated both data processing and analysis. This report reveals that the signs of any anomalous signals systematically diminish as the initial selection criteria are relaxed for degrouping while subsequent filtering remains stringent, leading to the complete elimination of any apparent non-Newtonian motion for binary separations within 0.5 pc. The study shows that any observed anomalous behaviour in solar neighbourhood wide binaries within 130 pc must be produced either by faint companion stars orbiting primary or secondary stars, or by flyby stars. The findings emphasize the importance of the choice of selection criteria in disentangling genuine binary dynamics from external influences. The conclusions align with the predictions of Newtonian mechanics and general relativity, though they do not exclude other phenomena at larger scales.

**Key words:** gravitation – methods: statistical – astrometry – binaries: general – stars: kinematics and dynamics – solar neighbourhood.

## 1 INTRODUCTION

Flat galactic rotation curves have been the subject of much study for over 50 yr (e.g. Rubin & Ford 1970; Roberts & Rots 1973; Rubin et al. 1978; Faber & Gallagher 1979). A number of explanations have been proposed to explain the phenomenon, most prominently dark matter (e.g. Ostriker & Peebles 1973; Bosma 1981a,b; Lequeux 1983). Dark matter is a postulated – but so far not observed – form of matter that would account for the failure of the velocity to fall off in the predicted manner by adding extra, unseen, mass to the equation. Another explanation is Milgromian Dynamics (MOND; e.g. Milgrom 1983; Sanders 1996; McGaugh & de Blok 1998). MOND theorizes that gravity falls off at a lesser rate when the gravitational field falls below a threshold value known as $a_0$ (Famaey & McGaugh 2012; Banik & Zhao 2022). This would also account for the observed flattening, which appears to persist out to as far as 1 Mpc if weak lensing is used to estimate the gravity (Brouwer et al. 2021; Mistele et al. 2024).

Because MOND operates in weak gravitational regimes, it has been hypothesized that this rotational flattening might also be apparent in binary star systems where the stars are separated widely enough that the acceleration scale is of a similar magnitude to the acceleration scale in spiral galaxies, i.e. about $1.2 \times 10^{-10}$ m s$^{-2}$. If observed in wide binary systems (WBs), it could serve as a test to distinguish between the MOND and dark matter theories. Possible examples of this anomaly have been shown, for example, by Hernandez, Jiménez & Allen (2012), Scarpa et al. (2017), Pittordis & Sutherland (2018), Hernandez et al. (2019), Pittordis & Sutherland (2019), Hernandez, Cookson & Cortés (2021), Chae (2022, 2023), Hernandez (2023), and Pittordis & Sutherland (2023).

⋆ E-mail: stephen.cookson@sca-uk.com

These wide binary results have been the subject of debate regarding the origin of the signal, with some authors suggesting that faint tertiary components might be the cause (Belokurov et al. 2020; Clarke 2020), or that they are simply miscategorized flybys, which would be more evident at wider separations (Pittordis & Sutherland 2023).

In this paper, three catalogues of candidate WBs are created from *Gaia* DR3 with different selection criteria, referred to here as C10, C14, and C25. Each catalogue differs only in the maximum allowed parallax uncertainty, varying from 1 per cent to better than 2.5 per cent. Each catalogue is sorted and where a star appears in more than one pair, any and all matching pairs are removed from the catalogue. This process is referred to in this paper as degrouping.

Fig. 1 shows the distribution of 1D velocities in the sky plane against separation of candidate WBs from the first catalogue, C10,

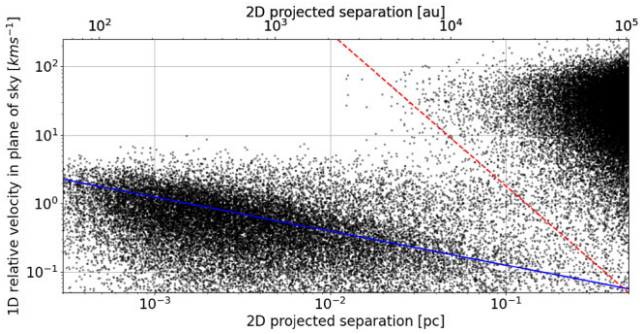

**Figure 1.** This plot shows catalogue C10, unfiltered and with separation out to 0.5 pc and with parallax uncertainty better than 1 per cent. The cloud at the lower left appears to be mostly genuine WBs, while the upper right cloud is mostly chance alignments. The blue line shows the expected Newtonian RMS prediction from Jiang & Tremaine (2010). The red dashed line shows the approximate separation between bound stars and apparent flybys, so a binary at this point has a 50 per cent probability of being bound or unbound.

after degrouping but before filtering. The figure shows two clusters: in the first cluster, in the middle of the figure, bound systems with stars in orbit around one another seem to be scattered around the blue Newtonian line. For the second cluster, at the top right, unbound stars (flybys) are scattered in a Gaussian pattern with a mean 2D velocity of about $38\sqrt{2}$ km s$^{-1}$ in this sample, calculated by deselecting those below the red dashed line. Because stars are only observed at a moment in time, due to the long orbital period, and we know neither the eccentricity of the orbit, nor its inclination to our sky plane, the data points do not fall exactly on the Newtonian line. Looking carefully at the region above the Newtonian line at a separation of ≈0.1 pc, the eye perceives a slight overdensity with a mean velocity of about 0.5 km s$^{-1}$, where the two groups overlap. This would appear to give rise to previous findings of non-Newtonian motion (e.g. Hernandez et al. 2012, 2019, 2021; Chae 2022, 2023; Hernandez 2023).

This study attempts to answer the question: do the criteria for inclusion or exclusion of certain candidate WBs affect the evidence of apparent non-Newtonian motion? Of particular interest are the treatments for low-intensity third stars and for flybys which may or may not be partners in a gravitationally bound binary pair.

The paper is structured as follows:

(i) Introduction. This section.

(ii) Methods. This section describes the download of the astrometry data on 1.5 million local stars (star profiles) and the further downloads of three catalogues of potential binary pairs of varying parallax precision. It also describes the processing of these catalogues into high-quality samples for analysis.

(iii) Data analysis. The results are presented in two tables and four charts, indicating that for each high-quality catalogue as the input noise threshold is relaxed, the number of 'clean' WBs is reduced, leading to a reduction in the apparent non-Newtonian signal.

(iv) Discussion. The discussion section discusses how the various methodological and data challenges were overcome.

(v) Conclusions. This section shows the conclusion that the relaxation of the initial selection criteria in the degrouping process increases the number of selected binaries at the beginning of the process. However, this leads to a reduction in the number of clean pairs post-degrouping and the total elimination of any apparent non-Newtonian effect.

## 2 METHODS

### 2.1 Method overview

In designing a methodological approach to assess the evidence for non-Newtonian dynamics, there are a number of objectives to take into account. They are to:

(i) identify and eliminate, or compensate for, close binaries (CBs);
(ii) identify and eliminate, or compensate for, flybys;
(iii) make optimal use of the quality assurance work already in place from Gaia Collaboration (2023);
(iv) produce a clear illustration of any non-Newtonian motion which may be found.

CBs are typically characterized by a low-quality signal and even missing values. As such they require a low signal accuracy threshold to allow them to be included in the sample. Other parameters, such as the 2D projected interstellar distance ($s_{2D}$) and the relative 1D velocities in the sky plane ($v_\alpha$ and $v_\delta$) of the WB, require the opposite: high accuracy data. The efforts of successive authors to improve the data quality through rigorous input noise filtering have potentially had the unintended effect of removing CBs from their considerations, meaning that apparently well-defined WBs could actually be triples with undetected CBs. Other authors have performed a second scan to reintroduce such CBs as a separate step in the process, or to quantify the impact of CBs through statistical modelling.

Meeting both these contradictory requirements simultaneously is one of the main challenges for this paper. However, the work of the *Gaia* team comes to our aid. The process for producing radial velocities (RVs) with the *Gaia* radial velocity spectrometer (RVS) has been an onerous task, diligently performed. With *Gaia* DR2 (Gaia Collaboration 2018), some 7 million stars brighter than 12 mag were matched with RVs. This was expanded to some 33 million stars in DR3 (Gaia Collaboration 2023) with magnitude brighter than 14.5. Not only are these relatively bright, high-quality stars, but where a star was found to be itself a binary, no RV was assigned. So for the purposes of this paper, the act of assigning an RV is a quality cut that further removes triple stars, though inevitably some remain.

The outline of the method is as follows:

(i) Download the full astrometric solutions, along with other selected attributes (star profiles) for some 1.5 million local stars.

(ii) Download three catalogues of potential binary pairs (C10, C14, and C25) of varying parallax uncertainty (better than 1 per cent, 1.4 per cent, and 2.5 per cent, respectively). An uncertainty limit of 2.5 per cent was found to allow many more CBs to be included in the selection for degrouping purposes than an uncertainty of 1 per cent, for instance.

(iii) Degroup. Where any star appears in more than one pair, all such pairs are rejected. High-noise CBs are used at this point to ensure that remaining WBs have as few CBs as possible.

(iv) Exclude WB pairs where either star has RV = null. This ensures that the majority of high-velocity flybys can later be eliminated. Any remaining stellar solutions which do not meet the *Gaia* team's RV quality threshold are also dropped at this stage, leading to a high-quality sample for the next step.

(v) Calculate 1D velocities in RA and Dec. with spherical projection corrections and 2D separations.

(vi) Apply further quality filters, e.g. RUWE[1] (to remove unseen CBs), distance (better resolution of CBs), and relative RV (to remove fast flybys).

(vii) Apply a Hertzsprung–Russell (HR) filter to remove more CBs.

(viii) Apply final relative 1D velocity filters for flybys and plot the kinematics against the expected Newtonian RMS prediction from Jiang & Tremaine (2010), assuming both stars follow the mass function from Kroupa, Tout & Gilmore (1993).

## 2.2 Data acquisition

Gaia Collaboration (2023) provides a new range of observations with more detailed parameters than previous releases. The *Gaia* DR3 data set provides a range of observations with detailed RVs, RUWE, FLAME[2] mass, age of star, binary probability, and error correlation values. The *Gaia* DR3 data set was downloaded using custom-built Python (Python Core Team 2019) software called BYNARY (https://github.com/SteveBz/Bynary). The new DR3 attributes allowed new features to be built into the software to take advantage of them. The greater precision of DR3 meant that the question of hidden third stars became a more subtle question not only around remaining hidden tertiaries, but also around many still dim, but now detectable, tertiary stars.

As in Hernandez et al. (2021), the download of binary pairs was broken into two parts, each part taking significant time. Part one of the download was concerned with selecting all possible individual star profiles within a specified distance, initially 143 pc, or a parallax of seven milliarcseconds. Part two, the pairings of stars close to each other, gives candidate binary systems. It is also time consuming to run and is designed so that part two can be run for multiple selections without part one needing to be rerun.

Because of the volume of stars (the first download includes 1.5 million solutions), data handling needs to be considered carefully. Some techniques are better performed on the ESA server, some on a local SQL data base, and still others are better performed in computer memory. The method described here attempts to make best use of the technologies available to process the data most efficiently.

### 2.2.1 Download of individual star attributes

The initial download was performed with only minimal concern to signal accuracy. This ensured that however many binary catalogues were later downloaded, star data would only need to be downloaded from the *Gaia* server once. It also helped to keep the underlying data model clean and normalized, which sped up queries and calculations. The SQL query published by El Badry (El-Badry, Rix & Heintz 2021) was modified for this purpose. It also had two further features that needed to be modified for use here. First, the query assumes Newtonian dynamics and filters out non-Newtonian binaries. Secondly the original query utilizes quality checks of the form '> nn' such that nulls always fail, e.g.

*and phot_rp_mean_flux_over_error* > 10

*and phot_bp_mean_flux_over_error* > 10

As some CB companions have null values for $RP$ and $BP$, they would be ignored by the query, resulting in binaries that are actually triple stars. BYNARY has incorporated a version of the El Badry query, modified in these ways, to be executed remotely on the ESA server and the results downloaded automatically.

For each degree in RA, stars are downloaded where:

(i) parallax $\geq$ 7 mas and < 1000 mas;
(ii) parallax_over_error > 5;
(iii) phot_g_mean_flux_over_error > 5.

### 2.2.2 Download three catalogues of candidate binary pairs

As already noted, some stars in the *Gaia* DR3 data set had $RP$ or $BP$ with null values. In order that these should be included in the degrouping exclusion criteria, when downloading *Gaia* candidate binaries, the colour–magnitude selections were considerably relaxed, with no selection on the $RP$ or $BP$ signal-to-noise ratios, and signal-to-noise ratio of >5 for *phot_g_mean_flux_over_error*. The three catalogues with parallax uncertainty of 1 per cent, 1.4 per cent, and 2.5 per cent were then selected (C10, C14, and C25, respectively). C25 allows more potential faint companion stars through the selection process for the purposes of deselecting potential triple systems. A distance of up to 143 pc for the binary pairs was chosen to speed up the download and also to allow for loss of granularity above 130 pc (as noted in Hernandez et al. 2021).

Pairing is best performed on the ESA server. The server runs ADQL, an astronomical version of SQL, with specialist commands such as POINT and CIRCLE. These commands allow the *Gaia* data base to be queried for close-by pairs, which could turn out to be binary star systems. The sky is divided by Gaia Collaboration (2023) into *n* HEALpixes.[3] The allowed values of *n* are 192 (level 2), 768 (level 3), 3072 (level 4), or 12 288 (level 5). Larger values (higher levels) are quicker and less error-prone on the ESA server. Level 2 can be quite slow and error-prone.

As selected binaries do not cross HEALpix boundaries, a balance has to be struck between (slow, unresponsive, and error-prone) and (fast, responsive, and error-free), but with fewer identified binary candidates due to edge effects. Here a value of 3072 was used for *n*. Stars within 0.5 pc of one another are selected as potential binaries. Nulls are treated as in Section 2.2.1.

## 2.3 Degrouping to exclude triple systems including CBs

The binary star pairs are sorted and arranged so that the brighter star is classified as the main star and the dimmer as its companion.

In addition to the missing $RP$ or $BP$ values, some stars also have a higher uncertainty around parallax_over_error. A star may of course have both a high parallax uncertainty and missing $RP$ or $BP$ values. As they all exert a gravitational pull, they should be included in the degrouping process, even if they are later excluded for having no RV.

The *Gaia* satellite has specialist photometric filters. $RP$ and $BP$ are the red and blue filters, respectively, with different naming

---

[1]RUWE is an acronym for Renormalized Unit Weight Error. It is expected to be around 1.0 for sources where the single-star model provides a good fit to the astrometric observations. A value significantly greater than 1.0 could indicate that the source is non-single or otherwise problematic for the astrometric solution. The accepted cut-off for RUWE is 1.25 (Andrew et al. 2022; Penoyre, Belokurov & Evans 2022).

[2]FLAME (Final Luminosity Age Mass Estimator) *Gaia* work package (Pichon 2007).

[3]HEALPix is an acronym for Hierarchical Equal Area isoLatitude Pixelation of a sphere. It is a way to divide a spherical surface so that every subdivision (HEALPix) covers the same solid angle.

**Table 1.** This table shows two examples of star systems that appear to be binary when the null $BP\_RP$ star is excluded by a $phot\_rp\_mean\_flux\_over\_error > 0$ condition (null is not greater than zero).

| *Gaia* ID | $BP - RP$ | $E(BP - RP)$ |
|---|---|---|
| Gaia DR3 845964280770594304 | 0.894566 | 0.2272 |
| Gaia DR3 845964285065622016 | *Null* | *Null* |
| Gaia DR3 845964285066533120 | 0.962605 | 0.0575 |
| Gaia DR3 918931175219973760 | 3.142 | 0.3723 |
| Gaia DR3 918939730796058880 | 3.395 | *Null* |
| Gaia DR3 918931175221235456 | *Null* | *Null* |

**Table 2.** This table shows examples of two star systems that appear to be binaries when the low quality star is excluded by pre-filtering.

| *Gaia* ID | $\varpi$ [mas] | $\sigma_\varpi$ [per cent] | $\sigma_\varpi$ [mas] |
|---|---|---|---|
| Gaia DR3 963335807006167552 | 12.8894 | 0.11 | 0.0140 |
| Gaia DR3 963337147035963392 | 12.9333 | 0.09 | 0.0123 |
| Gaia DR3 963305536076389632 | 12.4595 | 1.18 | 0.1468 |
| Gaia DR3 1068492931584022016 | 9.0094 | 0.13 | 0.0119 |
| Gaia DR3 1068494439116192000 | 8.9570 | 0.16 | 0.0145 |
| Gaia DR3 1068492931584554752 | 9.3412 | 2.17 | 0.2022 |

conventions. An SQL query on this star of the type:

$phot\_bp\_mean\_flux/phot\_bp\_mean\_flux\_error > 10$

for a CB with a null $BP$ would only give a single pair and the dim companion star would drop out of the data set, implying, falsely, that the main star and its companion were true binaries. Examples of the effect of including a star with a null $RP$ or $BP$, or a low-quality star are shown in Tables 1 and 2, respectively.

After downloading the pairs of stars (the candidate binaries), in order to exclude triple systems, all potential solutions are selected, including those with high noise ratios or missing parameters. These pairs are concatenated into a single list of primaries and secondaries and sorted. Where any star appeared in more than one pair, all such matching pairs were rejected. Therefore, the remaining pairs all had unique stars which did not appear in more than one pair. Stars in different binaries were separated by at least the minimum separation, in this case 0.5 pc. Degrouping is performed on the local data base.

### 2.4 RV check

To facilitate the removal of flybys in future steps, RVs and proper motions are needed. For any pairs of stars where either star lacks a measured RV (i.e. where RV = null), the pair was deselected. As already mentioned, *Gaia* is quite selective over which stars have RVs. In DR3, stars with a magnitude dimmer than 14.5, missing astrometric parameters, potential spectroscopic binaries, or large error do not have assigned RVs. By definition, this step also excludes low accuracy stars, so no further signal quality checks are required. RV checking is performed on the local data base.

### 2.5 Calculate 1D velocities and 2D separations

Remaining pairs were then imported into computer memory for further processing. It is at this point that 1D velocities are calculated from the proper motions from *Gaia* and spherical corrections are performed as in Hernandez et al. (2021), shown below.

The first step is to calculate the absolute velocity of the primary star in X, Y, and Z (Smart 1968, pp 16–21).[4]

Individual stars are placed at the mean parallax distance and the individual parallaxes are replaced by the inverse variance weighted mean binary parallax $\overline{\varpi}$ (see Banik 2019; Banik et al. 2024).

$$\overline{\varpi} = \frac{\varpi_1 \sigma_{\varpi_2}^2 + \varpi_2 \sigma_{\varpi_1}^2}{\sigma_{\varpi_1}^2 + \sigma_{\varpi_2}^2}, \quad (1)$$

$$U_1 = -\kappa \frac{\mu_{\alpha 1}^*}{\overline{\varpi}} \sin(\alpha_1) - \kappa \frac{\mu_{\delta 1}}{\overline{\varpi}} \sin(\delta_1) \cos(\alpha_1)$$
$$+ v_{r1} \cos(\delta_1) \cos(\alpha_1), \quad (2)$$

$$V_1 = \kappa \frac{\mu_{\alpha 1}^*}{\overline{\varpi}} \cos(\alpha_1) - \kappa \frac{\mu_{\delta 1}}{\overline{\varpi}} \sin(\delta_1) \sin(\alpha_1)$$
$$+ v_{r1} \sin(\alpha_1) \cos(\delta_1), \quad (3)$$

$$W_1 = \kappa \frac{\mu_{\delta 1}}{\overline{\varpi}} \cos(\delta_1) + v_{r1} \sin(\delta_1). \quad (4)$$

For units, unless otherwise stated, distances are in pc, parallaxes in mas, proper motions in mas yr$^{-1}$, and velocities in km s$^{-1}$.[5]

Now calculate the spherical correction to projection effects for velocities in the sky plane (El-Badry 2018) for the primary star only:

$$\mu_{\alpha s} = \mu_{\alpha 1}^* - \frac{\overline{\varpi}}{\kappa} \left[ -U_1 \sin(\alpha_2) + V_1 \cos(\alpha_2) \right], \quad (5)$$

$$\mu_{\delta s} = \mu_{\delta 1} - \frac{\overline{\varpi}}{\kappa} [-U_1 \cos(\alpha_2) \sin(\delta_2)$$
$$+ V_1 sin(\alpha_2) \sin(\delta_2) + W_1 \cos(\delta_2)]. \quad (6)$$

Calculate relative 1D velocities between the main and companion stars in the sky plane:

$$v_\alpha = \frac{\kappa}{\overline{\varpi}} |\mu_{\alpha 2}^* - \mu_{\alpha 1}^* + \mu_{\alpha s}|, \quad (7)$$

$$v_\delta = \frac{\kappa}{\overline{\varpi}} |\mu_{\delta 2} - \mu_{\delta 1} + \mu_{\delta s}|. \quad (8)$$

Calculate 2D separation in the sky plane:[6]

$$\cos(\theta) = \sin(\delta 2) \sin(\delta 1) + \cos(\delta 2) \cos(\delta 1) \cos(\alpha_2 - \alpha_1). \quad (9)$$

The 2D projected distance between the binary stars ($pc$) is

$$s_{2D} = \frac{d_1 \sigma_{\varpi_2}^2 + d_2 \sigma_{\varpi_1}^2}{\sigma_{\varpi_1}^2 + \sigma_{\varpi_2}^2} \sqrt{2(1 - \cos(\theta))}. \quad (10)$$

The resulting collection of candidate WBs is presented in Fig. 2, where the effects of degrouping can be seen in the plane of the Milky Way around (0, 0).

---

[4] $[n]_1$: any symbol with a subscript 1 is a value pertaining to the primary star.
$[n]_2$: any symbol with a subscript 2 is a value pertaining to the secondary, or companion, star.
[5] $U_1$, $V_1$, and $W_1$: velocity in km s$^{-1}$ of primary star in equatorial Cartesian coordinates.
$\kappa$: 4.740 470 464. Conversion factor between au yr$^{-1}$ and km s$^{-1}$.
$\alpha_1$: Right ascension (RA) of primary star, in decimal degrees.
$\delta_1$: Declination (Dec.) of primary star, in decimal degrees.
$\mu_{\alpha 1}^*$: proper motion in RA (pmra) for primary star in mas yr$^{-1}$. The * superscript indicates values multiplied by $\cos(\delta_1)$, which is how *Gaia* provides it.
$\varpi_1$, $\varpi_2$, $\overline{\varpi}$: parallax of primary and secondary stars, and their inverse variance weighted mean in milliseconds of arc (mas).
$v_r$: line-of-sight recession velocity in km s$^{-1}$.
[6] $\theta$: angle between stars from the Solar system barycentre. Only $\cos(\theta)$ is relevant. $d_1, d_2, d$: distance to star 1, star 2, and the inverse variance weighted mean distance ($pc$), respectively.

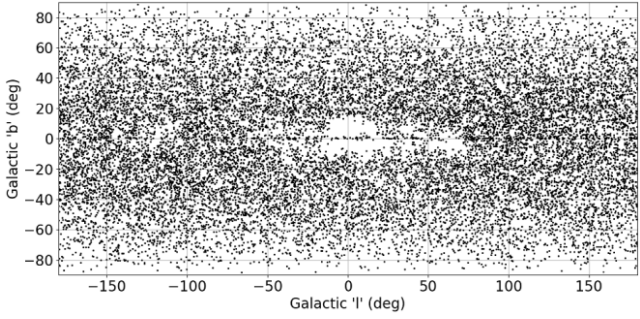

**Figure 2.** The C25 wide binary catalogue is shown post-degrouping in Galactic coordinates. The effects of degrouping can be seen around the Galactic centre and along the Galactic equator.

### 2.6 Error propagation

The relative velocity $v$ along each direction is[7]

$$v = \frac{\kappa}{\varpi}(\mu_1 - \mu_2). \tag{11}$$

The uncertainties in $\mu_1$, $\mu_2$, and $\varpi$ are denoted by $\sigma_{\mu_1}$, $\sigma_{\mu_2}$, and $\sigma_{\varpi}$, respectively. Systematic errors in the astrometry are neglected because these small-angle effects should cancel when taking the relative velocity between the stars in each WB given that it covers only a small part of the sky (Vasiliev 2019).

The full error propagation is given in Appendix A.

### 2.7 Applying quality filters

Filters were applied that distinguished between the different groups or helped to accentuate them. An RUWE of <1.25 was applied (which helps to deselect potential third stars). In addition, a maximum RV difference cut of 4 km s$^{-1}$ was initially applied as the mean, absolute 1D velocity of flybys is 38 km s$^{-1}$ in this sample, so cutting at 4 km s$^{-1}$ should remove most random associations. A distance cut-off of 130 pc was used (distant close companions are harder to detect) as this was found to be most effective. Stars for which *Gaia* has provided the RV and three magnitudes ($RP$, $G$, and $BP$) were those stars that the *Gaia* Team had already determined to have high-quality solutions. Additional filters seemed to reduce the number of stars by a small amount without visibly affecting the shape or parameters of the kinematic plot (Section 2.9). For this reason, most plots rely on the choices of Gaia Collaboration (2023). No additional quality cuts have been applied at this stage in order to show the greatest number of binaries.

### 2.8 HR cut

The HR diagram provides another opportunity to remove close companions. For main-sequence stars, a pair of stars which are indistinguishable and close will appear spuriously brighter than a single star at the same temperature. This gives rise to the grey parallel track on the HR diagram which is visible about 0.75 mag above the selected main sequence between an $RP$ - $BP$ of 0.75 and 2 (as also evident in fig. 2 of Banik et al. 2024). BYNARY interpolates around a parallelogram (parallel to the main sequence and sheared with respect to the absolute *Gaia*-band magnitude) between about 0.70 and 2 in $RP$ - $BP$ and 4.7 and 8.7 in absolute magnitude, deselecting stars outside a bar 0.8 mag high. The full height of the HR strip between 1 and 2 is about 1.4–1.6 mag. 0.8 is about half of this and ensures that the upper half of the diagram (about 0.75 less than the lower half) is always deselected. If the 0.8 value is expanded, the final kinematic plot rapidly deteriorates. Deselected stars are shown in light grey in Fig. 3.

---

[7] $\mu_1$ and $\mu_2$ are the proper motions of the two stars in RA or Dec.

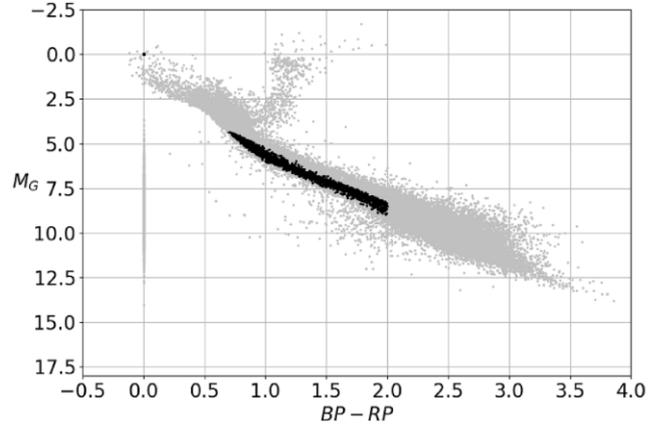

**Figure 3.** Catalogue C14 after applying HR filters. Selected binaries are shown in black and deselected ones in light grey.

This HR cut was found to be critically effective at removing triples masquerading as binaries (see also Belokurov et al. 2020). The light grey vertical bar at $RP$ - $BP$ = 0 is caused by stars with no $RP$ or $BP$. In this sample, 332 pairs have one or more stars with $RP$ - $BP$ = 0. These are also deselected.

### 2.9 Kinematic plot

Finally, in the Kinematic plots of 1D velocity against the projected separation (see Section 3), three further filters were applied. A separation cut of 0.1 pc and a relative 1D velocity filter of <4 km s$^{-1}$ was applied to remove flybys, which also motivates the cut in RV difference. Moreover, the fastest 2 per cent of WBs were ignored to remove any remaining pairs that might still contain CBs that had not been detected yet (the *Gaia* binary probability parameter has an average of 0.13 across all the samples, so 2 per cent is conservative). Finally, any velocities to the right of the red 50 per cent probability line on Fig. 1 are excluded for being a flyby.

The WBs are then binned. The vertical value of each bin, $v_\alpha$ or $v_\delta$, is the RMS value of the velocities in the bin along the east–west or north–south direction. For each bin (either RA or Dec.), the vertical error bars, $E(\sigma_{bin})$, corrected for sample size, are given as the quadrature sum of $\sigma_{bin}$ and the uncertainty in the independent variables; in this case the 1D velocity in either the RA or Dec. direction ($\sigma_{v_\alpha}$ or $\sigma_{v_\delta}$):

$$\sigma_{v_\alpha}^2 = \frac{\sum \sigma_{v_\alpha}^2}{n}, \tag{12}$$

$$E(\sigma_{bin}) = \sqrt{\frac{\sigma_{bin}^2}{2n} + \sigma_{v_\alpha}^2}, \tag{13}$$

with a similar approach used for the Dec. direction. The uncertainty estimate in $E(\sigma_{bin})$ includes contributions from both measurement uncertainty and intrinsic velocity dispersion. Since the observed velocity dispersion is partly due to measurement errors, this approach provides a conservative estimate of the total uncertainty. Additionally, the plotted $x$-value for each bin is the geometric mean of the projected separations in that bin.

**Table 3.** This table shows the WB-counts for each stage in the cleansing process by catalogue. The columns show the catalogue name, the number of pairs initially downloaded, the number of pairs after degrouping and with a known RV for both stars, the number of pairs filtered by RUWE, 1D velocity and distance, those which pass the HR cut, and the number of pairs in the final kinematic plot. The last two columns are the estimated critical velocity ($v_c$) at which any non-Newtonian kinematics becomes apparent, together with any estimated critical separation ($s_c$). ∗ indicates that the value has been augmented by relaxing the RV and relative velocity cut-off to 10 km s$^{-1}$ and the separation cut-off out to 0.5 pc, allowing closer and more distant companions, respectively (the wider RV cut also allows for a more uncertain RV).

| Catalogue | Download [#] | Degrp + RV [#] | Qual Cut [#] | HR Cut [#] | Kin. Plot [#] | $v_c$ [km s$^{-1}$] | $s_c$ [pc] |
|---|---|---|---|---|---|---|---|
| XH22 | - | - | 2k | 444 | 423 | 0.45 | 0.01 |
| C10 | 88k | 21.6k | 5.4k | 1057 | 999 | 0.27 | 0.03 |
| C14 | 121k | 21.7k | 5.2k | 1013 | 959 | 0.25 | 0.03 |
| C25 | 180k | 21.4k | 4.9k | 941 | 891 | - | - |
| C25∗ | 180k | 21.4k | 6.2k∗ | 991∗ | 902∗ | - | - |

**Table 4.** This table shows how the quality of the *Gaia* parallax varies as a mean percentage uncertainty for each catalogue. The columns show the mean parallax uncertainty, $\mathbb{E}[\sigma_\varpi/\varpi]$, at each stage, though the final column (Kin. Plot 2) shows the median value.

| Catalogue | Download [per cent] | Degrp + RV [per cent] | Qual Cut [per cent] | HR Cut [per cent] | Kin. Plot 1 [per cent] | Kin. Plot 2 [per cent] |
|---|---|---|---|---|---|---|
| C10 | 0.41 | 0.27 | 0.14 | 0.12 | 0.12 | 0.14 |
| C14 | 0.53 | 0.30 | 0.14 | 0.12 | 0.12 | 0.14 |
| C25 | 0.77 | 0.35 | 0.13 | 0.12 | 0.12 | 0.14 |

## 3 DATA ANALYSIS

The results are summarized in Tables 3 and 4. Regardless of the starting quality, each catalogue, after degrouping and with an assigned RV for both stars, has a mean Parallax/Error ≈300–400 and ends with mean Parallax/Error >800.

As will be seen, the process of relaxing the initial quality cut (i.e. injecting about 100k faint noisy sources into the process) allows more systems to be included at intermediate stages of the pipeline. None of these have an associated RV and will just disappear again. However, about 100 will combine with apparently clean binary pairs, allowing more CBs to be removed at the degrouping stage.

The analysis starts with an initial cut-off at 0.1 pc where most unbound stars are to the right. Hernandez et al. (2021), with 423 selected pairs from *Gaia* Early DR3 (Gaia Collaboration 2021), identified a potential non-Newtonian region starting at a critical separation of about $s_c = 0.01$ pc, i.e. $s_c > 0.01$ pc and below a 1D velocity $v_c \approx 0.4$–0.5 km s$^{-1}$. This corresponds well to the tilt of the apparent overdensity in Fig. 1 relative to the blue Newtonian line.

Each plot is divided into six bins in RA 1D relative motion (red dashed error bars) and five bins in Dec. 1D relative motion (green dashed error bars), giving 11 bins in all. The number of WBs in each bin is shown as a label above the bin. The vertical value of the bin line is $v_\alpha$ or $v_\delta$, the RMS 1D velocity of the stars in the bin. Because the plot density is lower at the edges of the diagram, the vertical error bar is shown at the geometric mean value of the projected separations for the bin. The magnitude of the vertical error bar is calculated using equations (12) and (13).

One material point is that the velocity and separation calculations and the error propagation have been improved upon since those in Hernandez et al. (2021). As mentioned in equation (1), individual stars are here placed at the mean parallax distance, calculated from the inverse variance weighted mean binary parallax $\overline{\varpi}$ (see Banik 2019; Banik et al. 2024). This update has resulted in a significant reduction of the apparent non-Newtonian signal.

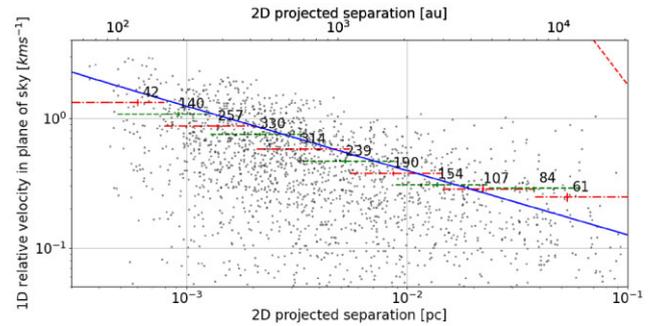

**Figure 4.** Catalogue C10, with an initial parallax uncertainty better than 1 per cent. The blue line represents the expected RMS Keplerian value from Jiang & Tremaine (2010) for a modelled distribution of binary masses. The red and green dot–dashed error bars represent the binned 1D relative velocity in RA and Dec, respectively. The numbers above the bins show the number of WBs in each bin. Note there are two points for each WB. The red dashed line to the top right is the ≈50 per cent flyby line from Fig. 1.

The solid blue line is the RMS Newtonian line taken from Jiang & Tremaine (2010), assuming both stars follow the mass function from Kroupa et al. (1993). The masses of the two stars in each binary are assumed to be independent.

Examining the first DR3 kinematic plot for catalogue C10 (initial parallax error for the download of <1 per cent) comprising 999 WBs (see Fig. 4), the apparent non-Newtonian region now starts at a separation of about 0.03 pc, triple the value quoted in Hernandez et al. (2021), and thus close to the predicted MOND radius $\sqrt{GM_\odot/a_0} = 0.034$ pc for a solar mass object. The critical velocity has a lower value of $v_c \approx 0.3$ km s$^{-1}$. After degrouping and other filtering, the mean parallax uncertainty of the plotted WBs is only 0.12 per cent.

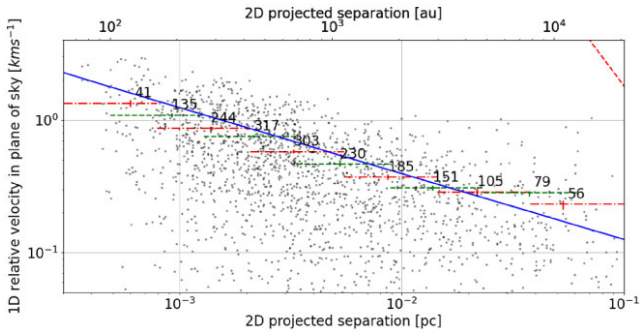

**Figure 5.** Similar to Fig. 4, but for catalogue C14, whose initial parallax error is <1.4 per cent.

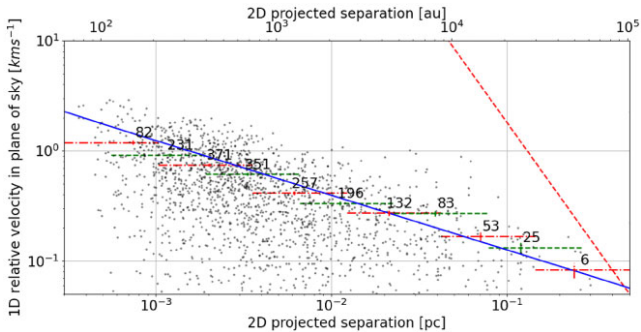

**Figure 6.** Similar to Fig. 4, but for catalogue C25, with an initial parallax uncertainty better than 2.5 per cent and separation now out to 0.5 pc. The axis ranges are extended to accommodate the wider range of separations.

The next catalogue analysed, labelled C14, had a slightly worse initial parallax uncertainty (<1.4 per cent) and ended up with 959 WBs (see Fig. 5). The apparent non-Newtonian region for this plot hardly changes, with $s_c \approx 0.03$ pc and $v_c \approx 0.25$ km s$^{-1}$. However, after degrouping and other filtering, the mean uncertainty of the plotted WBs is unchanged at 0.12 per cent, showing that the relaxing of the input quality has not adversely affected the quality of the output.

Finally, the C25 sample with still fewer systems (891 pairs), and an initial parallax uncertainty still worse (<2.5 per cent) shows no sign of non-Newtonian behaviour even out to $s_{2d} = 0.5$ pc (see Fig. 6). Degrouping and filtering keep the mean uncertainty at 0.12 per cent, again showing that the further relaxing of the input quality has not adversely affected the quality of the output. It actually has the opposite effect by helping to remove more systems with undetected CBs.

## 4 DISCUSSION

One difficulty in the use of binaries lies in selecting appropriate star systems to include in the analysis, i.e. identifying true binaries. The higher quality of the *Gaia* DR3 data set permits a greater degree of precision in identifying bound versus unbound systems and tertiary systems.

Separating bound stars from unbound stars was the first problem. Because there is an $\alpha$ and a $\delta$ 1D velocity for each WB, each has two data points (not the primary and secondary). It may be that as a non-compliant data point (e.g. the $v_\alpha$ data point) is removed for violating a quality cut, it takes a second compliant data point with it

(e.g. $v_\delta$). It is beyond the scope of this analysis to combine $v_\alpha$ and $v_\delta$ into a single quantity $v_{sky}$, as done in some works (Banik 2019; Banik et al. 2024). The approach taken in this study is more akin to Hernandez et al. (2023)

Fig. 1 presents the entire population of candidate binary stars for catalogue C10 after degrouping. It shows two clusters of secondary stars. These seem to correspond to bound pairs and unbound pairs, with no clear demarcation between them. The analysis starts with an initial cut-off at 0.1 pc (see Fig. 4), where most unbound stars are to the right (however, the final plots went out to 0.5 pc; see Fig. 6 with no difference in results). Initially, the focus is on the selection of those stars to the left of 0.1 pc and with a 1D relative velocity <4 km s$^{-1}$. For this reason, an RV difference cut of 4 km s$^{-1}$ is also applied. The sky-projected mean 2D velocity of the flyby stars in the sample is $38\sqrt{2}$ km s$^{-1}$ (see Fig. 1), so cutting at 4 km s$^{-1}$ removes most random associations/chance alignments. Even so, clearly there must be some overlap: bound companion stars (maybe those with heavier primaries) will appear in the cloud of predominantly unbound flybys on the right. Similarly, flybys will appear in the cloud of bound secondaries in the centre of the plot (because they are paired with lighter apparent 'primary' stars). A candidate secondary on the dividing line could equally well be a real pairing or just a flyby, based on the mass of the primary and the angle of the orbit to the sky plane.

One of the new attributes available in DR3 is binary probability.[8] These had a mean of 13 per cent on all three of WB catalogues, indicating that potentially up to 13 per cent of individual stars could actually be binaries themselves, making the selected candidate binary in fact a triple. For this reason, in each separation bin in all DR3 plots, the WBs in the top 2 per cent of relative velocity were removed. In the more widely separated bins, this could also be said to have allowed for unidentified flybys and chance alignments.

Table 3 shows the number of binary pairs at each stage for maximum parallax uncertainties at download time of 1 per cent, 1.4 per cent, and 2.5 per cent. The critical velocity $v_c$, at which non-Newtonian behaviour becomes apparent, reduces extract-by-extract, and the critical separation $s_c$ increases, until they disappear off the plot on the last row of the table. At the same time, the final mean parallax uncertainty of the binaries plotted remains at 0.12 per cent.

Eliminating all stars to the right of the 50 per cent flyby line has no effect for any of Figs 4–6, indicating that most if not all potential flybys have been successfully removed.

## 5 CONCLUSION

The gradual expansion of the degrouping process by relaxing the initial selection criteria progressively allows the removal of more pairs of stars where one appears to be associated with a low-quality, third, close star. As shown in Tables 3 and 4, as the initial maximum uncertainty requirement is relaxed from 1 per cent to 1.4 per cent and again to 2.5 per cent, the number of binaries selected increases from 88k, via 121k to 180k. However, this consequently leads to more candidate binaries being dropped by the degrouping process. After degrouping, the number of clean pairs drops from 21.6k to 21.4k, even though the pre-degrouping count more than doubles. At the same time, the final mean parallax accuracy of the binaries plotted remains 0.12 per cent. Additionally, most very faint stars (those dimmer than 14.5 mag) will not have been assigned an RV by the *Gaia* team – pairs will be dropped for this reason too.

---

[8]classprob_dsc_combmod_binarystar: probability from DSC-Combmod of being a binary star *Gaia* Documentation.

Further CBs are removed by the distance filter, the RUWE filter, and the HR cut, leaving only 'true' binaries and flybys. Flybys are removed by relative 1D velocity filters (both in the sky plane and radially) at 4 km s$^{-1}$, the 50 per cent flyby cut-off filter, and the separation cuts at 0.1 or 0.5 pc. Any remaining flybys or unidentified triples are removed by removing the fastest 2 per cent of stars, which itself is well below the 13 per cent of stars which are binaries as estimated by the *Gaia* binary probability parameter.

To summarize, despite doubling the number of downloaded pairs, the actual number of WBs plotted drops by 11 per cent, i.e. from 999 binaries with an apparent non-Newtonian signal to 891 pairs with no such signal, and a 0.12 per cent parallax uncertainty. In other words, the inclusion of an extra 100k high noise pairs (those with with an uncertainty of 1 per cent to 2.5 per cent, some of which may include CBs) only reduces the number of solutions plotted. This seems to be because none of them had RVs, so they were all excluded by the 'RV not equal to null' check, but some of them included CBs and the degrouping process excluded any such associated primaries from the selection.

Taken together, the evidence presented here suggests that there is either an undetectably small or no anomalous kinematic effect in wide binary stars within the solar neighbourhood. This is in keeping with Newtonian mechanics and general relativity, but does not necessarily exclude MOND or dark matter as a possibility at galactic scales or greater. DM is not expected to be detectable at stellar distance scales. MOND postulates the external field effect (see Banik & Zhao 2018; Banik et al. 2024), meaning that as the stellar system is embedded in the broader Galactic (External) field, MOND is not expected to be easily measurable here either. Even so, the 20 per cent uplift to the Newtonian relative velocity mentioned in Hernandez (2023) and Banik et al. (2024) should be visible in the final chart (Fig. 6). No such anomaly is detected. This finding is in keeping with other authors (e.g. Pittordis & Sutherland 2023; Banik et al. 2024). It is also in keeping with the lack of observed MOND effects in the Solar system e.g. the findings of Desmond, Hees & Famaey (2024) on Cassini radio tracking of Saturn and those of Vokrouhlický, Nesvorný & Tremaine (2024) on the energy distribution of long-period comets and the orbital inclination distribution of trans-Neptunian objects.

### ACKNOWLEDGEMENTS

As an amateur astronomer, the author would like to express thanks for the support received from many people. First, Professor Xavier Hernandez, from the Mexican University of UNAM, who was the inspiration for this. Also to Dr Mike McCulloch from Plymouth University for his support. Emeritus Professors Rick Diz and David Rees tirelessly proofread this document – under their guidance it was rewritten at least twice. Zac Plumber wrote the *Gaia* access code. Members of my astronomy society have had to live with this project and have supported me strongly over the last year and more, and the Mid Kent Astronomical Society has also been supportive. My thanks also for the very positive support of Prof David Bacon from University of Portsmouth, who introduced me to the discipline and tools required for submitting a professional quality paper. Finally, I am very grateful to the MNRAS referees for their constructive reviews and support.

This work has used data from the European Space Agency (ESA) mission *Gaia* (https://www.cosmos.esa.int/gaia), processed by the *Gaia* Data Processing and Analysis Consortium (DPAC; https://www.cosmos.esa.int/web/gaia/dpac/consortium). Funding for the DPAC has been provided by national institutions, in particular the institutions participating in the *Gaia* Multilateral Agreement.

### DATA AVAILABILITY

All the data here were downloaded from the *Gaia* collaboration at https://gea.esac.esa.int/archive/. It needs a login.

The python code to BYNARY is available for download and use by interested parties at https://github.com/SteveBz/Bynary.

Currently the software has only been installed on the Ubuntu family of operating systems. Requests for support to install on other Linux operating systems or other operating systems would be welcomed.

Please email me with any questions.

## APPENDIX A: ERROR PROPAGATION FOR BINARY STAR SYSTEM – STEP-BY-STEP APPROACH

### A1 Relative velocity

The relative velocity $v$ along each 1D direction in the sky plane is

$$v = \frac{\kappa}{\varpi} \cdot (\mu_1 - \mu_2), \tag{A1}$$

where $\varpi$ is the inverse variance weighted mean:

$$\varpi = \frac{\sigma_{\varpi,2}^2 \cdot \varpi_1 + \sigma_{\varpi,1}^2 \cdot \varpi_2}{\sigma_{\varpi,1}^2 + \sigma_{\varpi,2}^2} \tag{A2}$$

### A2 Define the variables and their uncertainties

*Primary star or secondary star, respectively:*

(i) $\alpha_1$ or $\alpha_2$: Right ascension
(ii) $\delta_1$ or $\delta_2$: Declination
(iii) $\varpi_1$ or $\varpi_2$: Parallax
(iv) $\mu_{\alpha,1}$ or $\mu_{\alpha,2}$: Proper motion in right ascension
(v) $\mu_{\delta,1}$ or $\mu_{\delta,2}$: Proper motion in declination

*Variances (diagonal elements):*

(i) $\sigma_{\alpha_1}^2, \sigma_{\delta_1}^2, \sigma_{\varpi_1}^2, \sigma_{\mu_{\alpha,1}}^2, \sigma_{\mu_{\delta,1}}^2$
(ii) $\sigma_{\alpha_2}^2, \sigma_{\delta_2}^2, \sigma_{\varpi_2}^2, \sigma_{\mu_{\alpha,2}}^2, \sigma_{\mu_{\delta,2}}^2$

*Covariances within each star:*

(i) $\mathrm{Cov}(\alpha_1, \delta_1) = \rho_{\alpha_1 \delta_1} \sigma_{\alpha_1} \sigma_{\delta_1}$
(ii) $\mathrm{Cov}(\alpha_1, \varpi_1) = \rho_{\alpha_1 \varpi_1} \sigma_{\alpha_1} \sigma_{\varpi_1}$
(iii) $\mathrm{Cov}(\alpha_1, \mu_{\alpha,1}) = \rho_{\alpha_1 \mu_{\alpha,1}} \sigma_{\alpha_1} \sigma_{\mu_{\alpha,1}}$
(iv) $\mathrm{Cov}(\alpha_1, \mu_{\delta,1}) = \rho_{\alpha_1 \mu_{\delta,1}} \sigma_{\alpha_1} \sigma_{\mu_{\delta,1}}$
(v) $\mathrm{Cov}(\delta_1, \varpi_1) = \rho_{\delta_1 \varpi_1} \sigma_{\delta_1} \sigma_{\varpi_1}$
(vi) $\mathrm{Cov}(\delta_1, \mu_{\alpha,1}) = \rho_{\delta_1 \mu_{\alpha,1}} \sigma_{\delta_1} \sigma_{\mu_{\alpha,1}}$
(vii) $\mathrm{Cov}(\delta_1, \mu_{\delta,1}) = \rho_{\delta_1 \mu_{\delta,1}} \sigma_{\delta_1} \sigma_{\mu_{\delta,1}}$
(viii) $\mathrm{Cov}(\varpi_1, \mu_{\alpha,1}) = \rho_{\varpi_1 \mu_{\alpha,1}} \sigma_{\varpi_1} \sigma_{\mu_{\alpha,1}}$
(ix) $\mathrm{Cov}(\varpi_1, \mu_{\delta,1}) = \rho_{\varpi_1 \mu_{\delta,1}} \sigma_{\varpi_1} \sigma_{\mu_{\delta,1}}$
(x) $\mathrm{Cov}(\mu_{\alpha,1}, \mu_{\delta,1}) = \rho_{\mu_{\alpha,1} \mu_{\delta,1}} \sigma_{\mu_{\alpha,1}} \sigma_{\mu_{\delta,1}}$

Similarly for the secondary star's parameters. The cross-covariances between the two stars are all zero.

### A3 Construct the original covariance matrix

The covariance matrix for a binary star system is a $10 \times 10$ matrix since there are five parameters for each star.

$$C_{\mathrm{orig}} = \begin{pmatrix} C_{1,1} & C_{1,2} \\ C_{2,1} & C_{2,2} \end{pmatrix}, \tag{A3}$$

where each block represents the covariances between different parameters:

$C_{1,1} = $ Covariances of primary star parameters.
$C_{2,2} = $ Covariances of secondary star parameters.
$C_{1,2} = C_{2,1}^T$
  $= $ Cross-covariances between primary and secondary star parameters.
    The cross-covariances between primary and secondary star parameters are assumed to be zero.

$$C_{\mathrm{orig}} = \begin{pmatrix} C_{1,1} & 0 \\ 0 & C_{2,2} \end{pmatrix} \tag{A4}$$

Then we can calculate the individual $5 \times 5$ matrices, but we compress into a $3 \times 3$ matrix for $C_{1,1}$ and $C_{2,2}$ because $v_\alpha$ does not depend on $\alpha$ or $\delta$ (and the same for $v_\delta$). The original covariant matrix ($C_{\text{orig}}$) becomes

$$
\begin{pmatrix}
\sigma_{\varpi,1}^2 & \rho_{\varpi\mu_{\alpha,1}}\sigma_{\varpi,1}\sigma_{\mu_{\alpha,1}} & \rho_{\varpi\mu_{\delta,1}}\sigma_{\varpi,1}\sigma_{\mu_{\delta,1}} & 0 & 0 & 0 \\
\rho_{\varpi\mu_{\alpha,1}}\sigma_{\varpi,1}\sigma_{\mu_{\alpha,1}} & \sigma_{\mu_{\alpha,1}}^2 & \rho_{\mu_\alpha\mu_{\delta,1}}\sigma_{\mu_{\alpha,1}}\sigma_{\mu_{\delta,1}} & 0 & 0 & 0 \\
\rho_{\varpi\mu_{\delta,1}}\sigma_{\varpi,1}\sigma_{\mu_{\delta,1}} & \rho_{\mu_\alpha\mu_{\delta,1}}\sigma_{\mu_{\alpha,1}}\sigma_{\mu_{\delta,1}} & \sigma_{\mu_{\delta,1}}^2 & 0 & 0 & 0 \\
0 & 0 & 0 & \sigma_{\varpi,2}^2 & \rho_{\varpi\mu_{\alpha,2}}\sigma_{\varpi,2}\sigma_{\mu_{\alpha,2}} & \rho_{\varpi\mu_{\delta,2}}\sigma_{\varpi,2}\sigma_{\mu_{\delta,2}} \\
0 & 0 & 0 & \rho_{\varpi\mu_{\alpha,2}}\sigma_{\varpi,2}\sigma_{\mu_{\alpha,2}} & \sigma_{\mu_{\alpha,2}}^2 & \rho_{\mu_\alpha\mu_{\delta,2}}\sigma_{\mu_{\alpha,2}}\sigma_{\mu_{\delta,2}} \\
0 & 0 & 0 & \rho_{\varpi\mu_{\delta,2}}\sigma_{\varpi,2}\sigma_{\mu_{\delta,2}} & \rho_{\mu_\alpha\mu_{\delta,2}}\sigma_{\mu_{\alpha,2}}\sigma_{\mu_{\delta,2}} & \sigma_{\mu_{\delta,2}}^2
\end{pmatrix} \tag{A5}
$$

**A4 Construct the Jacobian matrix and Jacobian transpose**

Creating the full Jacobian $J$:

$$
J = \begin{pmatrix}
\frac{\partial v_\alpha}{\partial \varpi 1} & \frac{\partial v_\alpha}{\partial \mu_{\alpha 1}} & \frac{\partial v_\alpha}{\partial \mu_{\delta 1}} & \frac{\partial v_\alpha}{\partial \varpi 2} & \frac{\partial v_\alpha}{\partial \mu_{\alpha 2}} & \frac{\partial v_\alpha}{\partial \mu_{\delta 2}} \\
\frac{\partial v_\delta}{\partial \varpi 1} & \frac{\partial v_\delta}{\partial \mu_{\alpha 1}} & \frac{\partial v_\delta}{\partial \mu_{\delta 1}} & \frac{\partial v_\delta}{\partial \varpi 2} & \frac{\partial v_\delta}{\partial \mu_{\alpha 2}} & \frac{\partial v_\delta}{\partial \mu_{\delta 2}}
\end{pmatrix} \tag{A6}
$$

Combining this with equations (A1) and (A2), and differentiating, we get the Jacobian matrix $J$, shown below, again ignoring the columns for RA and Dec.:

$$
J = \begin{pmatrix}
-\frac{v_\alpha}{\varpi}\varpi_1' & \frac{\kappa}{\varpi} & 0 & \frac{v_\alpha}{\varpi}\varpi_2' & -\frac{\kappa}{\varpi} & 0 \\
-\frac{v_\delta}{\varpi}\varpi_1' & 0 & \frac{\kappa}{\varpi} & \frac{v_\delta}{\varpi}\varpi_2' & 0 & -\frac{\kappa}{\varpi}
\end{pmatrix} \tag{A7}
$$

and the transpose $J^T$ is

$$
J^T = \begin{pmatrix}
-\frac{v_\alpha}{\varpi}\varpi_1' & -\frac{v_\delta}{\varpi}\varpi_1' \\
\frac{\kappa}{\varpi} & 0 \\
0 & \frac{\kappa}{\varpi} \\
\frac{v_\alpha}{\varpi}\varpi_2' & \frac{v_\delta}{\varpi}\varpi_2' \\
-\frac{\kappa}{\varpi} & 0 \\
0 & -\frac{\kappa}{\varpi}
\end{pmatrix} \tag{A8}
$$

where $\varpi_1'$ and $\varpi_2'$ are given by

$$\varpi_1' = \frac{d\varpi}{d\varpi 1} = \frac{\sigma_{\varpi,2}^2}{\sigma_{\varpi,1}^2 + \sigma_{\varpi,2}^2} \tag{A9}$$

$$\varpi_2' = \frac{d\varpi}{d\varpi 2} = \frac{\sigma_{\varpi,1}^2}{\sigma_{\varpi,1}^2 + \sigma_{\varpi,2}^2} \tag{A10}$$

**A5 Transform $C_{orig}$ using**

$$C_v = J \cdot C_{\text{orig}} \cdot J^T \tag{A11}$$

*A5.1 Multiply elements and gather terms for $v_\alpha$ and $v_\delta$*

$C_v$ is the covariance matrix for the transformed variables $v_\alpha$ and $v_\delta$.

$$
\begin{aligned}
\sigma_{v\alpha}^2 = (C_v)_{11} &= J_{1k} C_{kj} J_{j1}^T \\
&= \frac{1}{\varpi^2} \cdot \left( v_\alpha^2 \cdot \left( \frac{\sigma_{\varpi,2}^2}{\sigma_{\varpi,1}^2 + \sigma_{\varpi,2}^2} \right)^2 \cdot \sigma_{\varpi,1}^2 - 2\kappa \cdot v_\alpha \cdot \frac{\sigma_{\varpi 2}^2}{\sigma_{\varpi 1}^2 + \sigma_{\varpi 2}^2} \cdot \rho_{\varpi_1,\mu_{\alpha 1}} \sigma_{\varpi 1} \sigma_{\mu_{\alpha 1}} \right. \\
&\quad \left. + \kappa^2 \cdot \sigma_{\mu_{\alpha,1}}^2 + v_\alpha^2 \cdot \left( \frac{\sigma_{\varpi,1}^2}{\sigma_{\varpi,1}^2 + \sigma_{\varpi,2}^2} \right)^2 \cdot \sigma_{\varpi,2}^2 - 2\kappa \cdot v_\alpha \cdot \frac{\sigma_{\varpi,1}^2}{\sigma_{\varpi,1}^2 + \sigma_{\varpi,2}^2} \cdot \rho_{\varpi\mu_{\alpha,2}} \sigma_{\varpi,2} \sigma_{\mu_{\alpha,2}} + \kappa^2 \cdot \sigma_{\mu_{\alpha,2}}^2 \right)
\end{aligned} \tag{A12}
$$

Then $\sigma_{v\delta}^2 = (C_v)_{22}$ can be calculated in a similar way for $J_{2k} C_{kj} J_{j2}^T$, but also just by replacing $\mu_\alpha$ by $\mu_\delta$ throughout.

This paper has been typeset from a TEX/LATEX file prepared by the author.